\documentclass{Rinton-P9x6}
\usepackage{epsfig}
\usepackage{rotating}
\begin{document}

\title{Temperature Dependence of the Casimir Force for Metals}

\author{I. Brevik and J. B. Aarseth}

\address{Department of Energy and Process Engineering, Norwegian 
University of Science and Technology, N-7491 Trondheim, Norway
\\E-mails: iver.h.brevik@mtf.ntnu.no; jan.b.aarseth@mtf.ntnu.no
}

\author{J. S. H{\O}ye}

\address{Department of Physics, Norwegian University of Science and Technology,
N-7491 Trondheim, Norway
\\
E-mail: johan.hoye@phys.ntnu.no} 

\author{K. A. Milton}

\address{Department of Physics and Astronomy, The University of Oklahoma, 
Norman,
 Oklahoma 73019 USA\\
 E-mail: milton@nhn.ou.edu}


\maketitle

\abstracts{Starting from the Lifshitz formula for the Casimir force between 
parallel plates  we calculate the difference between the forces at two 
different settings, one in which the temperature is $T_1=350$ K, the other 
when $T_2=300$ K. As material we choose gold, and make use of the Drude 
dispersion relation. Our results, which are shown graphically, should be 
directly comparable to experiment. As an analogous calculation based upon 
the plasma dispersion relation leads to a different theoretical force 
difference, an experiment of this kind would be a decisive test. We also 
present an analogous calculation for the case when the two plates are 
replaced with a sphere-plate system, still with gold as material in both 
bodies. The sphere is assumed so large that the proximity theorem holds. 
Discussion of the consistency with the third law of thermodynamics and the
validity of the surface impedance approach is provided.
}

\section{Introduction}
Recently, there has been something of a heated controversy concerning
the temperature dependence of the Casimir effect between parallel metallic
plates.  The classic result for the pressure on one ideal metal plate separated
from another by a distance $a$ is\cite{milton01} ($\beta=1/T$;
in this paper we set $\hbar=c=k_B=1$)
\begin{equation}
{\mathcal   F}^T
=-\frac{\pi^2}{240
a^4}\left[1+\frac{1}{3}\left(\frac{2a}{\beta}\right)^4\right]\quad
(aT\ll1).
\label{m1}
\end{equation} 
Recently, however, it has been suggested\cite{bostrom00} that the transverse
electric zero mode should not contribute, which if the metal is otherwise
regarded as ideal (reflection coefficients equal unity), would lead to
a presumably observable linear temperature  correction to
the Casimir force, and a violation of the third law of 
thermodynamics.\cite{bordag01,geyer}  However, real metals do not possess ideal
reflection coefficients, and so are more complicated, and there seems
to be no contradiction with the Nernst heat theorem.\cite{hoye03}
It is the purpose of this paper to address the observable and thermodynamical
consequences on the Casimir pressure
of using the observed permittivity of real metals, particularly
gold, which has been used in many recent experiments.

\section{Proof that TE Zero-Mode Does Not Contribute at  $T=0$}

The crucial observation that the transverse electric
zero mode does not contribute is based on the condition
$\lim_{\omega\to0}\omega^2\varepsilon(\omega)=0$.
This follows from the general dispersion relation\cite{schwinger98}
\begin{equation}
\chi(\omega)=\frac{\varepsilon(\omega)-1}{4\pi}=\frac{\omega_p^2}{4\pi}
\int_0^\infty
d\omega'\frac{p(\omega')}{\omega^{\prime2}-(\omega^2+i\epsilon)^2}.
\end{equation}
Here the spectral function is positive, $p(\omega)>0$, and satisfies
the sum rule\footnote{The normalization of $p(\omega)$ plays no role in
our argument.  The high-frequency behavior of the susceptibility given by
the Drude formula is accepted, for which the spectral function is
$p(\omega)=(2/\pi)\gamma/(\omega^2+\gamma^2)$.}
\begin{equation}
\int_0^\infty d\omega'\,p(\omega')=1.
\label{sumrule}
\end{equation}
The structure of this dispersion relation allows us to make the complex
frequency rotation,
$\omega\to i\zeta$,
so
\begin{equation}
\chi(i\zeta)=\frac{\omega_p^2}{4\pi}
\int_0^\infty d\omega'\frac{p(\omega')}{\omega^{\prime2}
+\zeta^2}.
\end{equation}
The proof  is now immediate, because
\begin{equation}
\zeta^2\int_0^\infty d\omega'\frac{p(\omega')}{\omega^{\prime2}
+\zeta^2}=1-\int_0^\infty d\omega'\,\frac{\omega^{\prime2}p(\omega')}
{\omega^{\prime2}
+\zeta^2},
\end{equation}
which uses the sum rule.  Since the last integral converges to
one as $\zeta\to0$, the desired limit is established.  
Of course, this
behavior is consistent with the Drude model, and not with the plasma model.

\section{Temperature Dependent Lifshitz Force}

We now recall the Lifshitz expression for the Casimir force between 
two parallel nonmagnetic plates,
characterized by a permittivity $\varepsilon(\omega)$,
and separated by a gap $a$. In the notation of 
Ref.~\cite{hoye03} the force per unit area can be written as
\begin{equation}
{\mathcal{F}}^T= -\frac{1}{\pi \beta a^3}{\sum_{m=0}^\infty}{}' 
\int_{m\gamma}^\infty y^2dy
\left[ \frac{A_me^{-2y}}{1-A_me^{-2y}}+\frac{B_me^{-2y}}{1-B_me^{-2y}} \right].
\label{1} 
\end{equation}
Here, $y$ is a dimensionless quantity, $y=qa$, with
\begin{equation}
 q=\sqrt{k_\perp^2+\zeta_m^2},\quad \zeta_m=2\pi m/ \beta,\quad \gamma=2\pi
  a/\beta,
\label{2}
\end{equation}
and $\bf{k_\perp}$ is the transverse wave vector (i.e., the component of 
$\bf k$ parallel to the plates). Further, we have
 defined the squared reflection coefficients $A_m$ and $B_m$ by
\begin{equation}
A_m=\left( \frac{\varepsilon p-s}{\varepsilon p+s}\right)^2,\quad B_m=
\left(\frac{s-p}{s+p}\right)^2,
\label{3}
\end{equation}
where the Lifshitz variables  $s$ and $p$ are given by
\begin{equation}
s=\sqrt{\varepsilon-1+p^2}, \quad p=q/\zeta_m.
\label{4}
\end{equation}
The permittivity $\varepsilon (i\zeta_m)$ is a function of the imaginary 
Matsubara frequency $\zeta_m$. If the medium is non-dispersive, $A_m(q)=A(p)$ 
and $ B_m(q)=B(p)$. The prime on the sum in Eq.~(\ref{1}) means that the $m=0$ 
term is counted with half weight.

The free energy $F$ per unit area follows from the relation 
${\mathcal{F}}^T=-\partial F/\partial a$:
\begin{equation}
 F=\frac{1}{2\pi \beta a^2}\sum_{m=0}^\infty{}' \int_{m\gamma}^\infty y\,dy 
 \left[\ln (1-A_me^{-2y})+\ln (1-B_me^{-2y})\right].
\label{5}
\end{equation}
Note that $A_m$ refers to the TM mode, $B_m$ refers to the TE mode.

We are looking for temperature effects in the Casimir force, and shall in the 
following focus attention on the following two temperatures, both easily 
accessible in the laboratory:
\begin{equation}
T_1=300 \,{\rm K},\quad T_2=350 \,{\rm K}.
\label{6}
\end{equation}
Specifically, we want to calculate the difference in Casimir pressure between 
the two temperatures:
\begin{equation}
\Delta {\mathcal F}= \mathcal{F}(350\,{\rm K})-\mathcal{F}(300\,{\rm K})=
|\mathcal{F}(300\,{\rm K})|-|\mathcal{F}(350\,{\rm K})|
\label{7}
\end{equation}
(i.e., for convenience we let $\Delta \mathcal{F}$ mean the difference between 
the magnitudes).
This quantity depends on the values of $A_m$ and $B_m$, which in turn depend 
on which dispersion relation is adopted for  $\varepsilon (i \zeta_m)$.  In 
this way we calculate a decisive quantity which in principle is directly 
comparable to experiment.

\section{ Dispersion Relation}

As in our preceding paper,\cite{hoye03} we chose gold as the material
of which the plates were composed. For this metal we have access to excellent 
numerical data for $\varepsilon(i \zeta)$ (courtesy of Astrid Lambrecht and 
Serge Reynaud). The data are shown graphically in 
Refs.~\cite{lambrecht00,hoye03}.
It turns out that for low and moderate frequencies, at least up to about 
$1.5\times 10^{15}$ rad/s (1 eV), the data are nicely reproduced by the Drude 
dispersion relation
\begin{equation}
\varepsilon(i\zeta)=1+\frac{\omega_p^2}{\zeta(\zeta+\nu)},
\label{8}
\end{equation}
where, at room temperature, the plasma frequency $\omega_p$ and the relaxation 
frequency $\nu$ are equal to
\begin{equation}
\omega_p=9.0\,{\rm eV}, \quad \nu=35\, {\rm meV}.
\label{9}
\end{equation}
Strictly speaking, one should take into account also the temperature dependence
of $\nu$, such that $\nu \rightarrow \nu(i\zeta, T)$:
\begin{equation}
\varepsilon(i\zeta, T)=1+\frac{\omega_p^2}{\zeta [\zeta+\nu(T)]}.
\label{10}
\end{equation}
Here, $\nu(T)$ can be calculated via use of the Bloch-Gr{\"u}neisen formula, 
as explained in Appendix C in Ref.~\cite{hoye03}.

At high frequencies, $\zeta > 2\times 10^{15}$ rad/s, the Drude formula gives
values for $\varepsilon$ which are too low.

In this context it is of interest to know: What frequency region gives the 
main contribution to the Casimir force? To analyze this point, it is convenient 
to go back to the expression (\ref{1}), from which it is seen that the most 
important region is when $y$ is of order unity, $y \sim 1$. Assuming that the 
transverse wave vector ${\bf k_\perp}$ does not dominate in the expression 
(\ref{2}) for $q$, we thus get the condition $2\pi m a/\beta \sim 1$, or
\begin{equation}
m\sim \frac{1}{2\pi}\frac{1}{aT}.
\label{11}
\end{equation}
If $a=1\; \mu$m, we have $aT=0.13$ at room temperature, resulting in $m=1$ as 
the dominant frequency mode. If $a=0.5\; \mu$m, we expect a somewhat smeared-out
 distribution over the lowest integer values for $m$. If $a=3\;\mu$m or higher, 
the $m=0$ mode should be highly dominant: the problem becomes a high-temperature
 problem. We have done explicit numerical calculations, reproduced here in 
Table 1 for convenience,\cite{hoye03} which confirm these expectations in 
detail. The numbers in the table are the relative percentage of each mode $m$, 
i.e., the quantities ${(\cal{F}}_m^T/{{\cal{F}}^T})\times 100$, where 
${\cal F}_m^T$ denotes the $m$th mode contribution to the force:
\begin{equation}
{\cal F}^T=\sum_{m=0}^\infty {\cal F}_m^T.
\label{12}
\end{equation}
The numbers in the table are calculated from the empirical results for 
$\varepsilon (i\zeta_m)$.
\begin{table}
\caption{Contribution from the various Matsubara frequencies for gold. 
The relative contribution of $\mathcal{F}^T$ is given, in percent, for each 
mode in the interval $m \in [0,7]$. Room temperature is assumed. Empirical 
permittivities inserted, for all frequencies.}
\centering
\begin{tabular}{|ccccccccc|} \hline
$a(\mu\textrm{m})$  & $m=0$  & $1$  & $2$  &  $3$  & $4$  & $5$  & 
$6$   & $7$    \\ \hline 
0.5              & 10.20  & 31.24  & 22.95  &  15.09  &  9.18  & 5.28   & 2.91 
   & 1.55 \\ 
1                & 20.07  & 49.37  & 20.83  & 6.97    & 2.03   & 0.54   & 0.14 
   & 0.03 \\ 
2                & 44.56  & 49.87  & 5.17   & 0.37    & 0.02  &&&\\
3                & 70.95  & 28.41  & 0.63   & 0.01 &&&&\\
4                & 88.88  & 11.07  & 0.05  &&&&&\\
5                & 96.58  & 3.42 &&&&&&\\
6                & 99.06  & 0.94 &&&&&&\\
7                & 99.76  & 0.24&&&&&&
\\ \hline
\end{tabular}
\end{table}

The Matsubara frequencies, in view of  Eq.~(\ref{2}), become
\begin{eqnarray}
\zeta_m=\left\{ \begin{array}{ll}
2.47 m\times 10^{14}\, {\rm rad/s}  \quad  {\rm for} \quad T=300\,{\rm K}, \\
2.88 m\times 10^{14}\, {\rm rad/s}  \quad  {\rm for}  \quad T=350\,{\rm K}.
\end{array}
\right.
\label{13}
\end{eqnarray} 
{}From this it follows that from $m=1$ up to about $m=6$, the frequencies stay 
so low that the Drude formula can be used with confidence. And from Table 1 we 
see that in particular for large distances, from $a=1\,\mu$m and upwards, 
these frequencies encompass the large majority of the mode contributions to 
the force. For small gap widths, $a=0.5\,\mu$m and lower, the situation may be 
more questionable as indicated by the first line in the table, but it seems 
that even in these cases we can use the Drude formula with sufficient accuracy 
to make a meaningful comparison with experiments, given the present 
experimental accuracy. In the following we will use the Drude formula 
throughout. This simplifies the calculation significantly.

\section{Calculated Results}

\begin{figure}[t]
%\figurebox{15pc}{15pc}{} % to have a box alone
\begin{center}
\epsfxsize=20pc % will enlarge or reduce the postscript figures based on the 
%xsize
\epsfbox{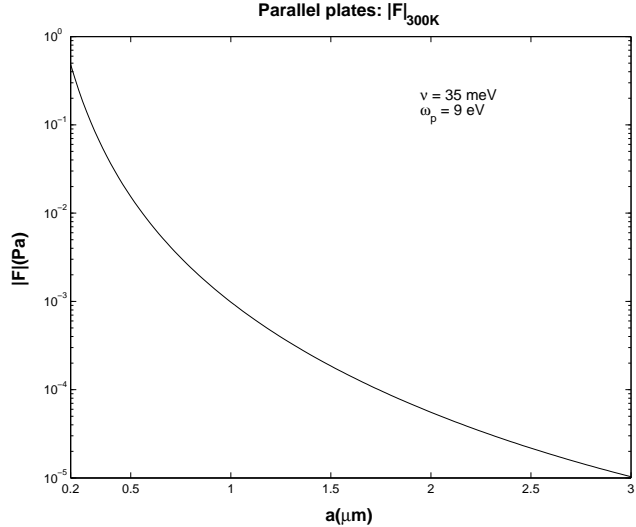} % postscript image file name
\end{center}
\caption{Magnitude of Casimir pressure between parallel plates, versus gap width 
$a$, when $T=300$ K. The curve for $T=350$ K overlaps the one shown. 
\label{fig:1}}
\end{figure}

Fig.~\ref{fig:1} shows how the Casimir force itself varies with $a$, for 
parallel 
gold plates. Since the force according to Eq.~(\ref{1}) is negative for 
attraction, we choose to present the magnitude $|\mathcal{F}^T|$ in the figure,
which for convenience is presented in semilog form.
 The curve is calculated for $T=300$~K, but a similar curve calculated for 
 $T=350$~K turns out to be visually indistinguishable from the curve shown.

\begin{figure}[t]
%\figurebox{15pc}{15pc}{} % to have a box alone
\begin{center}
\epsfxsize=20pc % will enlarge or reduce the postscript figures based on the 
%xsize
\epsfbox{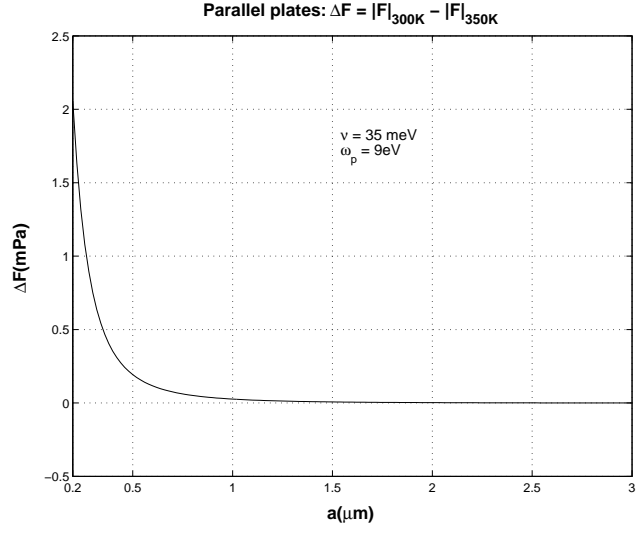} % postscript image file name
\end{center}
\caption{Force difference between parallel plates, Eq.~(\ref{7}), versus gap 
width. \label{fig:2}}
\end{figure}

Fig.~\ref{fig:2} shows the force difference $\Delta \mathcal{F}$, calculated 
according to Eq.~(\ref{7}). Typical orders of magnitude for 
$\Delta \mathcal{F}$ are in the millipascal range, 
when $a$ is less than about $0.5\,\mu$m. 
It is notable that it is the low-temperature term, $T=300$ K, which yields the 
strongest force at low and intermediate distances.  (When $a> 2.8\, \mu$m, 
$\Delta \mathcal{F}$ changes sign.) We note that the influence from temperature
in the formula (\ref{1}) is rather complex since $T$ occurs at three different 
places: (i) in the prefactor; (ii)  in the lower limit of the integral, and 
(iii) in the dependence of $A_m$ and $B_m$ on $T$ via the permittivity 
$\varepsilon=\varepsilon(i 2\pi mT)$. (We will discuss the temperature 
dependence of the relaxation frequence $\nu$ below.)
Actually we showed the essentials of the 
temperature dependence of the force already in Fig.~5 in Ref.~\cite{hoye03}: 
Assuming the Drude formula, and ignoring the temperature dependence of the
relaxation frequency, we found the force $|\mathcal{F}^T|$ to 
{\it diminish} with increasing $T$ up to $aT \simeq 0.35$. For higher values 
of $aT$, the force was found to increase again. In view of this, indeed we
expect that the lower-temperature term dominates in 
Fig.~\ref{fig:2} for low and intermediate distances. (Note also that the change
 in sign for $a=2.8\,\mu$m, mentioned above, corresponds nicely to the expected 
 transitional region since in this case $aT \simeq 0.37$.)

So far, we have assumed a constant relaxation frequency, $\nu=35$ meV. Will 
our results be changed significantly if we take into account the temperature 
dependence of $\nu$; cf. Eq.~(\ref{10})? The answer turns out to be no. We 
have made an explicit calculation of this, based upon the Bloch-Gr{\"u}neisen 
formula, yielding $\nu=35.6$ meV for $T=300$~K, and $\nu=41.8$ meV for 
$T=350$~K. The results were visually indistinguishable from those given in 
Fig.~\ref{fig:2}, so that our conclusion is that the assumed constancy of 
$\nu$ is justified for practical purposes.

\begin{figure}[t]
%\figurebox{15pc}{15pc}{} % to have a box alone
\begin{center}
\epsfxsize=20pc % will enlarge or reduce the postscript figures based on the 
%xsize
\epsfbox{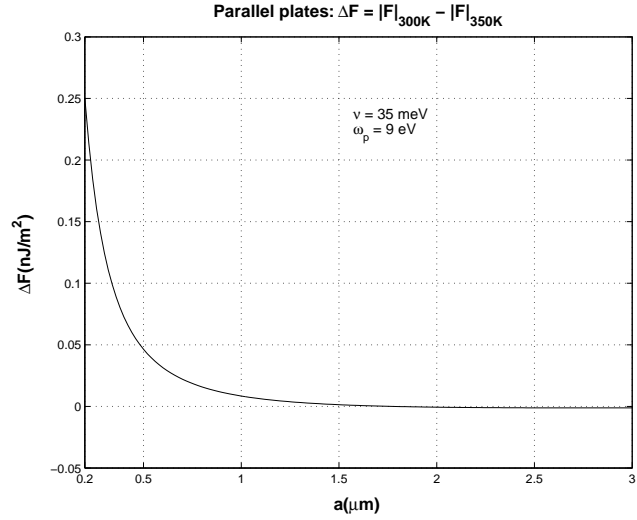} % postscript image file name
\end{center}
\caption{Free energy difference for parallel plates, Eq.~(\ref{14}), versus 
gap width. \label{fig:3}}
\end{figure}

Fig.~\ref{fig:3} shows how the corresponding difference in free energy, 
\begin{equation}
\Delta F = F(350\,{\rm K}) -F(300\,{\rm K})=|F(300\,{\rm K})|-|F(350\,{\rm K})|,
\label{14}
\end{equation}
varies versus $a$ for parallel plates. As expected, it is again the case 
$T=300$~K which is the dominant one for small and moderate gap widths.

\begin{figure}[t]
%\figurebox{15pc}{15pc}{} % to have a box alone
\begin{center}
\epsfxsize=20pc % will enlarge or reduce the postscript figures based on the 
%xsize
\epsfbox{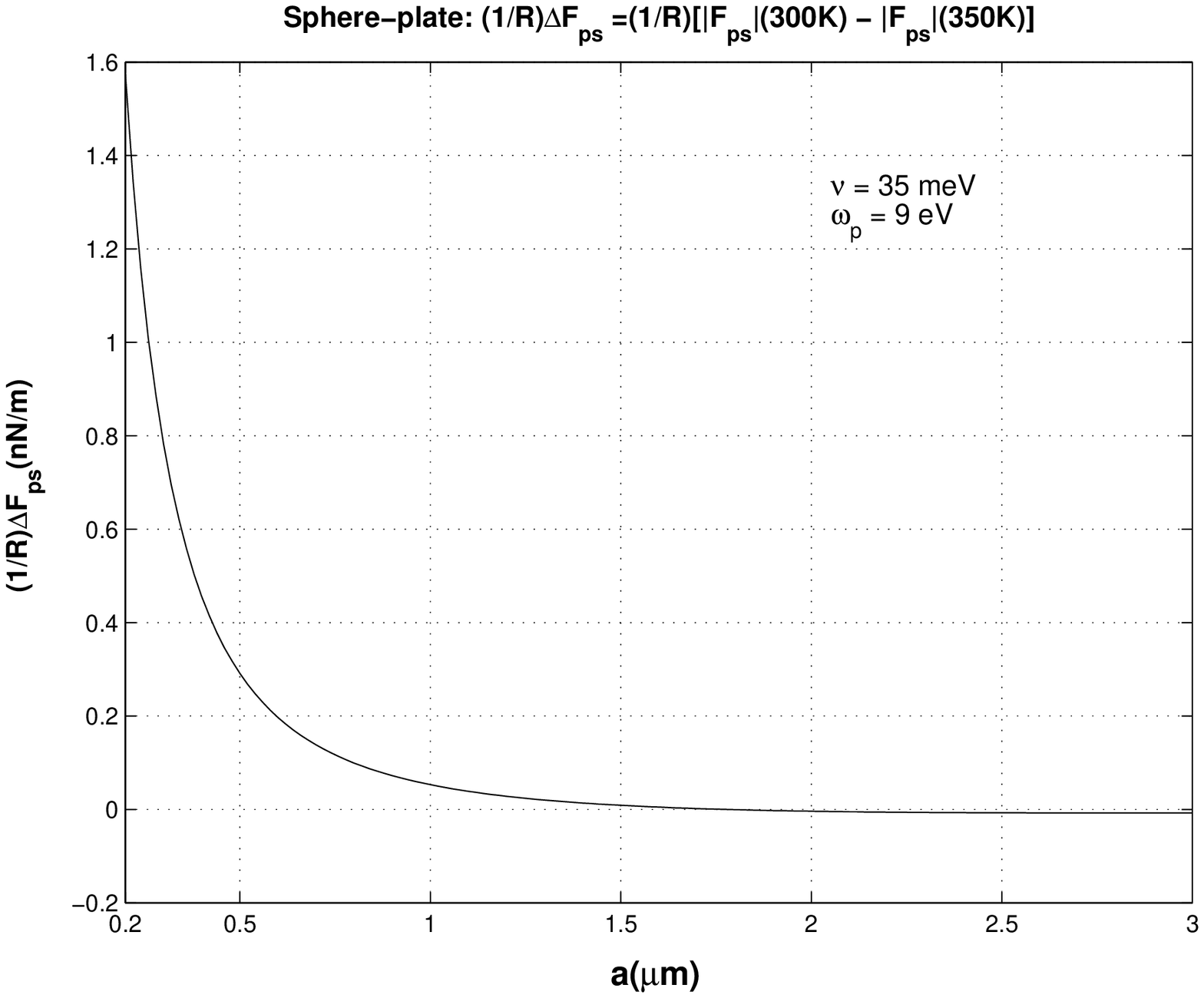} % postscript image file name
\end{center}
\caption {Force difference
 between sphere and plate, Eq.~(\ref{16}), versus gap width $a$.
  \label{fig:4}}
\end{figure}

Finally, we consider the case which is probably the one of principal 
experimental interest, namely a sphere of radius $R$ situated above a plane 
surface. Let the minimum distance between the spherical surface and the plane 
be $a$. Assuming the sphere large enough for the proximity theorem to be 
valid,\cite{milton01} we have for the force $\mathcal{F}_{\rm ps}^T$ on the 
sphere:
\begin{equation}
\mathcal{F}_{\rm ps}^T=2\pi R F(a),
\label{15}
\end{equation}
where $F(a)$ is the free energy for parallel plates as given by Eq.~(\ref{5}). 
Fig.~\ref{fig:4} shows how the difference
\begin{equation}
\frac{1}{R}\Delta \mathcal{F}_{\rm ps}=\frac{1}{R}\left[ \mathcal{F}_{ps}(350\,
{\rm K})-\mathcal{F}_{ps}(300\,{\rm K}) \right]
\label{16}
\end{equation}
varies with $a$. Again, it is the lower-temperature term that is the dominant 
one.

A dedicated experiment to look for the temperature dependence we have proposed
is probably essential to settle this issue.  The recent experiment by
Decca et al.\cite{decca03} is claimed to be in disagreement with our prediction.
However, that comparison is in fact not based on our detailed calculations,
and the experiment is subject to large, uncontrolled errors.\cite{iannuzzi03}

\section{Behavior of the Free Energy at Low Temperature}
\begin{figure}[t]
%\figurebox{15pc}{15pc}{} % to have a box alone
\begin{center}
\epsfxsize=20pc % will enlarge or reduce the postscript figures based on the 
%xsize
\begin{turn}{270}
\epsfbox{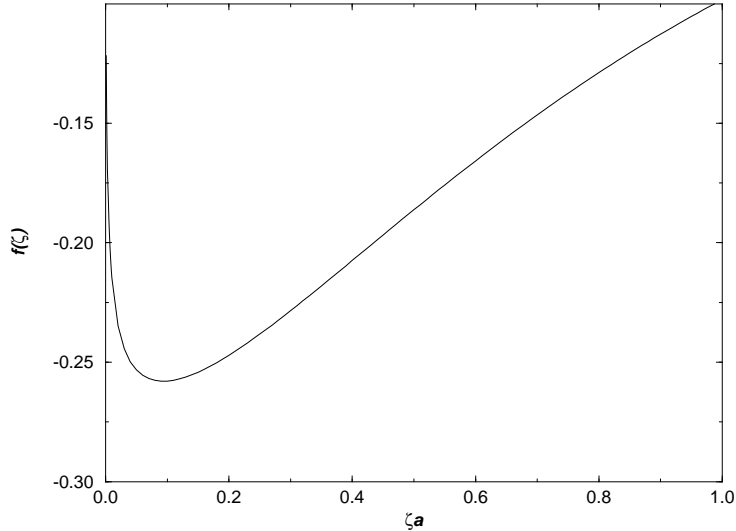} % postscript image file name
\end{turn}
\end{center}
\caption{The behavior of the free energy for low frequencies, in the Drude
model, with parameters suitable for gold, and a plate separation of
$a=1$ $\mu$m.  Here 
$ F^{\rm TE}=\frac{T}{2\pi a^2}\sum_{m=0}^\infty{}'f(\zeta_m)$.\label{fig:5}}
\end{figure}

The low temperature correction is dominated by low frequencies,\footnote{This
statement is in the context of using of the Euler-Maclaurin summation formula
to evaluate Eq.~(\ref{1}), for
example.} where the Drude formula is extremely accurate.  Using this fact, we
have performed analytic and numerical calculations which
show the free energy has a quadratic low-temperature dependence,
independent of the plate separation:
\begin{equation}
F(T)=F_0+T^2\frac{\omega_p^2}{48\nu}(2\ln2-1)=F_0+T^2(19\mbox{ eV}),
\label{quad}
\end{equation}
putting in the numbers for gold,
rather than the naive extrapolation
\begin{equation}
F=F_0+T\frac{\zeta(3)}{16\pi a^2}=F_0+\frac{T}{4\pi a^2}0.30
\end{equation}
We see from Fig.~\ref{fig:5} that this value indeed results if one extrapolates
the approximately linear curve there for $\zeta a>0.25$ to zero, following
the argument given in Eq.~(2.8) of Ref.~\cite{hoye03}.  However, we see
that the free energy smoothly changes to the quadratic behavior exhibited
in Eq.~(\ref{quad}).

Results consistent with these have been reported by Sernelius and 
Bostr\"om.\cite{sernelius}

\section{Surface impedance form of reflection coefficient}
It has been proposed that the resolution to the temperature problem
for the Casimir effect is that the surface impedance form of the reflection
coefficients should be used in the Lifshitz formula,\cite{geyer}
rather than that based on the bulk permittivity.  Here we show that the
two approaches are in fact equivalent, and that the former must include
transverse momentum dependence.

For the TE modes, the reflection coefficient is given by\cite{schwinger98} 
\begin{equation}
r^{\rm TE}=-\frac{k_{1z}-k_{2z}}{k_{1z}+k_{2z}},
\label{rte}
\end{equation}
where 
\begin{equation}
k_z=\sqrt{\omega^2\varepsilon-k_\perp^2}\to i\sqrt{\zeta^2[\varepsilon(i\zeta)-1]
+q^2},\label{disp}
\end{equation}
with $q^2=k_\perp^2+\zeta^2$,
and the subscripts 1 and 2 refer to the metal and the
vacuum regions, respectively.  Now from Maxwell's equations outside sources
we easily derive  just inside
the metal (the tangential components of $\bf E$ and $\bf B$ are continuous
across the interface)
\begin{eqnarray}
-ik_{1z}{\bf k_\perp\cdot B_\perp}-i\omega \varepsilon\left(1-\frac{k_\perp^2}
{\omega^2\varepsilon}\right){\bf k_\perp\cdot( n\times E_\perp)}&=&0,
\label{me1}
\\
-ik_{1z}{\bf k_\perp\cdot(n\times E_\perp)}-i\omega{\bf k_\perp\cdot B_\perp}
&=&0.\label{me2}
\end{eqnarray}
Here $\bf n$ is the normal to the interface. Now the surface impedance
is defined by
\begin{equation}
{\bf E_\perp}=Z(\omega,{\bf k_\perp}){\bf B_\perp\times n},\quad
{\bf n\times E_\perp}=Z(\omega,{\bf k_\perp}) {\bf B_\perp}.
\end{equation}
So eliminating $\bf B_\perp$ using this definition
we find two equations:
\begin{eqnarray}
k_{1z}=-\frac{\omega}Z,\label{kZ}\\
k_{1z}^2=\omega^2\varepsilon-k_\perp^2,
\end{eqnarray}
the latter being the expected dispersion relation (\ref{disp}).
Substituting this into the expression for the reflection coefficient
we find
\begin{equation}
r^{\rm TE}=-\frac{\zeta+Zq}{\zeta-Zq}=-\frac{1+Zp}{1-Zp},\quad p=\frac{q}{\zeta},
\end{equation}
which apart from (relative) signs (presumably just a different convention choice)
coincides with that given in Geyer et al.\cite{geyer} or Bezerra et 
al.\cite{bezerra}.

\subsection{Dependence on Transverse Momentum}

However, it is crucial to note that the ``surface impedance'' so defined
depends on the transverse momentum,
\begin{equation}
Z=-\frac\zeta{\sqrt{\zeta^2[\varepsilon(i\zeta)-1]+q^2}},
\end{equation}
and so $r^{\rm TE}\to 0$ as $\zeta\to 0$ just as in the dielectric constant
formulation.  Of course, we have exactly the same result 
for the energy as before, since
this is nothing but a slight change of notation.

It is therefore incorrect to assume 
that $Z$ is only a function of frequency, not of transverse momentum, and
to use the normal and anomalous skin effect formulas derived for real waves
impinging on imperfect conductors.\footnote{Of course, in general, the
permittivity will be a function both of the frequency and the transverse
momentum, $\varepsilon(\omega,\mathbf{k}_\perp)$, but we believe the latter
dependence
is not significant for separations larger that $\hbar c/\omega_p=0.02$ $\mu$m.} 

 How does the usual argument go?   The normal component of the
wavevector in a conductor is given by
\begin{equation}
k_z=\left[\omega^2\left(\varepsilon+i\frac{4\pi\sigma}\omega\right)-k_\perp^2
\right]^{1/2}\to\sqrt{i4\pi\omega\sigma},
\label{nse}
\end{equation}
from which the usual normal skin effect formula follows immediately,
\begin{equation}
Z(\omega)=-(1-i)\sqrt{\frac\omega{8\pi\sigma}}.
\end{equation}
However, the last step here consists in omitting two ``small''
terms: $\varepsilon$ (okay) and $k_\perp^2\le\omega^2$.  
Here this last is
not valid because in going to finite temperature we have severed the
connection between $\omega\to i\zeta$ and $k_\perp$; the latter is in no
sense ignorable as we take $\zeta\to0$ to determine the low temperature
dependence. This is the same error
to which we refer in  our published paper.\cite{hoye03}

These considerations are consistent with those of Esquivel et al.\cite{esquivel}
\section{Conclusions}

Our results of main interest are probably those shown in Figs.~\ref{fig:2} and 
\ref{fig:4}. These 
force curves show a dependence upon $a$ that reflect our underlying choice of 
the Drude dispersion relation. We can compare our results with those recently 
obtained by Chen et al.\cite{chen03}  They make use of the plasma 
dispersion relation instead of the Drude relation, and obtain results for the 
Casimir forces that differ from ours even in {\it sign}. The force curves thus 
give rise to a very useful critical test, in principle. It would be quite 
interesting if the experimentalists could measure these force curves directly.
We have also commented on the purported violation of basic principles of
thermodynamics and on the claimed necessity to use surface impedances at
$\sim1$ $\mu$m plate separations and find no merit in these objections.
\section*{Acknowledgments}
We thank Raul Esquivel for helpful conversations.  KAM thanks the US
Department of Energy for partial support of his research.

\end{document}